%% file: pascual_WCL2021-1674.R1.tex
\documentclass[journal,12pt,onecolumn]{IEEEtran}
\usepackage{cite}
\usepackage{amsmath,amssymb,amsfonts}
\usepackage{algorithmic}
\usepackage{graphicx}
\usepackage{url}
\usepackage{bbm}
\usepackage{bm}
\usepackage{siunitx}
\usepackage{mathtools, nccmath}
\graphicspath{ {Images/} }
\usepackage{textcomp}
\usepackage{xcolor}
\usepackage{mathtools}
\usepackage{blindtext}
\usepackage[utf8]{inputenc}
\usepackage{float}
\usepackage{enumitem}
\usepackage{hyperref}
\usepackage{multirow}
\usepackage{tabularx}
\usepackage{soul}

\newcommand{\prob}[1]{\mathbb{P}\left(#1\right)}  % Probability statement
\newcommand{\probc}[2]{\mathbb{P}\left(#1 \mid #2\right)}  % Conditional Probability
\newcommand{\esp}[1]{\mathbb{E}\left[#1\right]}

\usepackage[caption=false]{subfig}
\usepackage{tikz}
\usetikzlibrary{math}
\usetikzlibrary{calc}

\usepackage{acronym}
\acrodef{cdf}[CDF]{cumulative distribution function}
\acrodef{pdf}[PDF]{probability density function}
\acrodef{los}[LOS]{line-of-sight}
\acrodef{nlos}[NLOS]{non-line-of-sight}
\acrodef{ppp}[PPP]{Poisson point process}
\acrodef{bs}[BS]{base station}
\acrodef{mmw}[mmWave]{millimeter wave}
\acrodef{rv}[r.v.]{random variable}
\acrodef{map}[MAP]{maximum a posteriori probability}
\acrodef{knn}[$k$NN]{$k$-nearest neighbors}
\acrodef{kns}[$k$NS]{$k$-nearest segments}
\acrodef{gps}[GPS]{Global Positioning System}
\acrodef{gaussian}[PG]{Parzen-Gaussian window}
\acrodef{naive}[NB]{naive Bayes}

\def\BibTeX{{\rm B\kern-.05em{\sc i\kern-.025em b}\kern-.08em
    T\kern-.1667em\lower.7ex\hbox{E}\kern-.125emX}}

\newcommand{\reappendix}[1]{\hyperref[#1]{Appendix~\ref*{#1}}}

\usepackage[normalem]{ulem}
    
\begin{document}
\title{\LARGE{LOS/NLOS Estimators for mmWave Cellular Systems with Blockages}
\thanks{The work presented in this paper has been funded through the project ROUTE56 - PID2019-104945GB-I00 (funded by Agencia Estatal de Investigación, Ministerio de Ciencia e Innovación, MCIN / AEI / 10.13039/501100011033).}
\thanks{© 2022 IEEE.  Personal use of this material is permitted.  Permission from IEEE must be obtained for all other uses, in any current or future media, including reprinting/republishing this material for advertising or promotional purposes, creating new collective works, for resale or redistribution to servers or lists, or reuse of any copyrighted component of this work in other works.}
}

\author{\IEEEauthorblockN{Tomàs Ortega\IEEEauthorrefmark{1},
Antonio Pascual-Iserte\IEEEauthorrefmark{2}, \IEEEmembership{Senior Member, IEEE},
Olga Muñoz\IEEEauthorrefmark{3}, \IEEEmembership{Member, IEEE}}\\
\IEEEauthorblockA{Dept. of Signal Theory and Communications, Universitat Politècnica de Catalunya, Barcelona, Spain\\
Email: \IEEEauthorrefmark{1}tomas.ortega@upc.edu,
\IEEEauthorrefmark{2}antonio.pascual@upc.edu,
\IEEEauthorrefmark{3}olga.munoz@upc.edu}\vspace{-0.9cm}}

% \markboth{IEEE WIRELESS COMMUNICATIONS LETTERS, VOL. 11, NO. 1, JANUARY 2022}{ORTEGA \MakeLowercase{\textit{et al.}}: LOS/NLOS ESTIMATORS FOR mmWAVE CELLULAR SYSTEMS WITH BLOCKAGES}

\maketitle

\begin{abstract} 
Designers of \ac{mmw} cellular systems need to evaluate \ac{los} maps to provide good service to users in urban scenarios. In this letter, we derive estimators to obtain \ac{los} maps in scenarios with potential blocking elements. Applying previous stochastic geometry results, we formulate the optimal Bayesian estimator of the LOS map using a limited number of actual measurements at different locations. The computational cost of the optimal estimator is derived and is proven to be exponential in the number of available data points. An approximation is discussed, which brings the computational complexity from exponential to quasi-linear and allows the implementation of a practical estimator. Finally, we compare numerically the optimal estimator and the approximation with other estimators from the literature and also with an original heuristic estimator with good performance and low computational cost. For the comparison, both synthetic layouts and a real layout of Chicago have been used.

\end{abstract}


\input{Sections/introduction}
\input{Sections/model_formulation}

\input{Sections/bayesian_predictor}
\input{Sections/implementation}
\input{Sections/results_alternativo}
\vspace{-0.3cm}
\input{Sections/conclusion}

\vspace{-0.3cm}
\input{Sections/appendix}

\vspace{-0.08cm}
\bibliography{blocking_references}
\bibliographystyle{IEEEtran} 

\end{document}

%% file: Sections/introduction.tex
\section{Introduction}
\IEEEPARstart{D}{ue} to the capacity requirements of 5G, mobile communications networks aim to exploit the benefits of the \ac{mmw} bands from 30 GHz to 300 GHz \cite{5G_NR}. They present important differences with conventional mobile communications, such as their higher sensitivity to blocking elements, as buildings. Because of the significant impact of blockages on communication, having accurate blockage maps when deploying a high-frequency  \ac{bs} is of utmost importance for operators. These maps can be obtained through drive tests, although they are costly and time-consuming.

In the recent literature, deep learning (DL) methods have been proposed to predict \ac{mmw} blockages from sub-6GHz channels measurements \cite{9121328} or past adopted beamforming vectors \cite{8646438}. DL methods have the advantage of decreasing the computational load at the classification stage at the cost of increasing it at the training stage to learn the statistics on the underlying process from the data. However, DL methods cannot outperform a Bayesian classifier which is the optimal approach in terms of classification error when the statistics of the event to be predicted are known. 

In this letter, we provide new methods to compute estimates or predictions of line-of-sight (LOS)/non-LOS (NLOS) maps. Specifically, we rely on recent results for the statistics of LOS/NLOS events \cite{randomshape, relay_positioning} to derive a Bayesian estimator that uses a limited set of measurements and scenario parameters, e.g., building density and probability distribution of their sizes. Also, we propose several more straightforward approaches and benchmark their estimation accuracy and computational complexity with the Bayesian (i.e., optimal) estimator.

The information required by the proposed methods can be obtained with conventional mobile user terminals (UTs) endowed with \ac{gps} capabilities. The set of collaborating UTs determine if they have coverage or not and their \ac{gps} coordinates. Then, they will send this information to the node in charge of making the predictions once they are in coverage. Therefore, the proposed procedure does not require specific or costly equipment for the measurements.

The estimates of the LOS/NLOS maps obtained with the proposed methods can be useful to operators to indicate where to locate infrastructure even in regions where maps are not available, not accurate, or out of date, which is quite frequent as cities are alive-entities. Using stochastic propagation models where building densities are determined offline, without the need for high-fidelity environmental data, can be particularly interesting for low-cost deployments.

%% file: Sections/model_formulation.tex
\section{Model and Problem Formulation} \label{sec:model_and_problem_formulation}
In this section, we first describe our model for a single cell scenario with blockage elements. Then, we formulate the LOS estimation problem. We will consider a bounded region of the plane, with a \ac{bs} at the center and containing potentially blocking elements, e.g., buildings. For simplicity, we will take our cell region to be a disk of radius $R$ centered at the \ac{bs}.

\subsection{Blocking Elements}
The positions of the blocking elements will be drawn from a \ac{ppp} with spatial density $\lambda$ \cite{randomshape}. Given a fixed point in our region of study and the link between the \ac{bs} and that point, let $K$ be the Poisson random variable (r.v.) counting the number of blocking elements effectively blocking the link. Then, the probability of the event consisting in having a clear link (i.e., with \ac{los}), $\prob{LOS}$, is given by \cite{randomshape, relay_positioning}:
\begin{equation}\label{eq:prob_los}
	\prob{LOS} = \prob{K = 0} = e^{-\esp{K}}.
\end{equation} 

\begin{figure}[t] 
    \centering
	\includegraphics[width=.33\columnwidth]{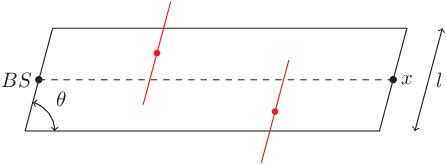}
    \caption{Parallelogram $S_{l\theta}$ between the \ac{bs} and the point of study $x$.}
    \label{fig:rectangle}
\end{figure}

The blocking elements will be modeled as segments with random lengths and orientations whose probability density functions (PDFs) are given by $f_L(l)$ and $f_{\Theta}(\theta)$, respectively. Accordingly, we will use upper-case $L$ (or $\Theta$)  to refer to the r.v., and lower-case $l$ (or $\theta$) to refer to a realization of that r.v. The expected value $\esp{K}$ is derived in \cite{randomshape}, and is given by
\begin{equation}
	\esp{K} = \int_l \int_{\theta} \lambda f_L(l) f_{\Theta}(\theta) A_{S_{l\theta}} \mathrm{d}l \mathrm{d}\theta,
\end{equation}
where $A_{S_{l\theta}}$ is the area\footnote{To account for the height of the buildings, we only need to modify the area $A_{S_{l\theta}}$ as in \cite{randomshape, relay_positioning}. However, there is no need if UTs are lower than buildings (e.g., if the users are persons in outdoor scenarios). In such a case, any building with any height between the BS and the UT will produce NLOS.} of the parallelogram $S_{l\theta}$ determined by the geometric locus where the center of the blocking elements with length $l$ and orientation $\theta$ must lie to block the given link (see \autoref{fig:rectangle}). More complex models could be used, but the simple line model already reflects reality quite well in urban scenarios where the blocking elements are buildings \cite{los_intervals}.

We can generalize the previous result to the case of multiple links as in \cite{relay_positioning}. Suppose that we are given a finite set of links, indexed by the set $\mathcal{J} \subset \mathbb{N}$. Let $K_{\mathcal{J}}$ be the Poisson r.v. counting the blocking elements that effectively block at least one of the links in $\mathcal{J}$. Then, the expected value of $K_{\mathcal{J}}$ is
\begin{equation}\label{eq:e_k_j}
	\esp{K_{\mathcal{J}}} = \int_l \int_{\theta} \lambda f_L(l) f_{\Theta}(\theta) A_{ \cup_{j \in \mathcal{J}}S_{j_{l\theta}} } \mathrm{d}l \mathrm{d}\theta,
\end{equation}
where $S_{j_{l\theta}}$ is the parallelogram defined above for the $j$th link. 

\subsection{Problem Formulation} 
\begin{figure}[t]
    \centering
    \includegraphics[width=.25\columnwidth]{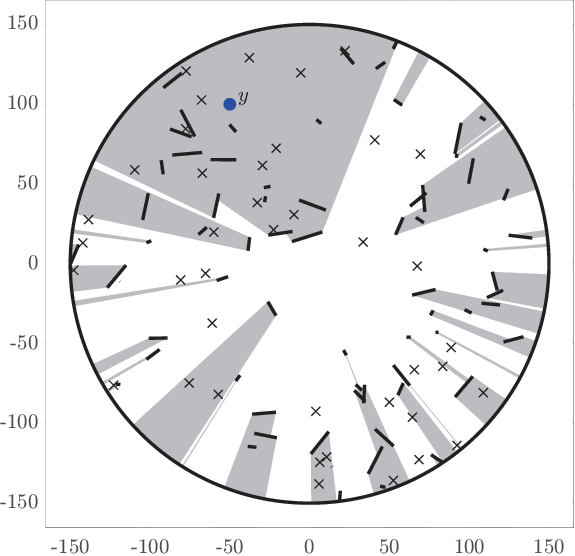}
    \caption{Example of scenario with $R = 150$ m. Black segments are obstacles, LOS regions are clear, NLOS regions are shaded, measurement points are crosses, and $y$ is the point where to estimate if it is in LOS or NLOS. 
    }
    \label{fig:scenario_example}
\end{figure}

Suppose now that we are given $N$ different sample points $\{x^{(i)}\}_{i = 1}^N$ in our region of study. Each $x^{(i)}$ will take a value $d^{(i)}$ which will be either $1$ (LOS) or $0$ (NLOS).  See \autoref{fig:scenario_example} for an example of our scenario.
Then, given any position $y$ in the region of study, we want to find a predictor $\hat{y}$ of its state.

%% file: Sections/bayesian_predictor.tex
\section{Bayesian Predictor} \label{sec:bayesian_predictor}

We will seek a predictor that minimizes the probability of miss-classification, that is, 
\begin{align*}
    \prob{\hat y \neq y} &=  \prob{\hat{y} = 1 \,|\, y = 0} \cdot \prob{y = 0} + \\
    & \phantom{{}={}} \prob{\hat{y} = 0 \,|\, y = 1}\cdot \prob{y = 1}.
\end{align*}
Such an estimator is given by the \ac{map} estimator\cite{kay}:
\begin{equation} \label{eq:map}
    \hat{y}_{\mathrm{MAP}}(\{x^{(i)}\}) = 
    \begin{cases}
        1 ~ \mathrm{if} ~  \probc{y=1}{\{x^{(i)} = d^{(i)}\}} > \\
        \hspace{1.1cm} \probc{ y = 0 }{ \{x^{(i)} = d^{(i)}\} },\\
        0 ~ \mathrm{otherwise},
    \end{cases}
\end{equation}
where we have clumped all the data in the conditional statement of the probability to avoid cumbersomeness. Since the two probabilities appearing above are complementary, that is,
\begin{equation*}
    \probc{ y=1 }{ \{x^{(i)} = d^{(i)}\} } = 1 - \probc{ y=0 }{ \{x^{(i)} = d^{(i)}\} },
\end{equation*}
we just have to find one of both terms, for example,
\begin{equation}\label{eq:y_conditioned_to_data}
    \probc{ y = 1 }{ \{x^{(i)} = d^{(i)}\} } = \frac{\prob{ y=1 , \{x^{(i)} = d^{(i)}\} }}{\prob{ \{x^{(i)} = d^{(i)}\} }}.
\end{equation}

The calculation of the numerator and the denominator in (\ref{eq:y_conditioned_to_data}) requires computing the probability that a set of $M$ points are in LOS or NLOS (with $M= N+1$ in the case of the numerator, and $M=N$ in the denominator). Assume,  without loss of generality, that the $M$ points are ordered so that the first $W$ points are in LOS, for some $0\leq W\leq M$, leading to 
\begin{multline}\label{eq:nlos_to_cond}
    \prob{ LOS_1, \dotsc , LOS_W, NLOS_{W+1}, \dotsc, NLOS_M } =\\
    {}\probc{  NLOS_{W+1}, \dotsc, NLOS_M }{ LOS_1, \dotsc , LOS_W } \cdot \prob{ LOS_1, \dotsc , LOS_W }.
\end{multline}
Expanding the last term yields (in the following, the symbol $\vee$ stands for ‘or’, while $\wedge$ stands for ‘and’):
\begin{align}\label{eq:nlos_cond_los}
    \probc{  NLOS_{W+1}, \dotsc, NLOS_M }{ LOS_1, \dotsc , LOS_W } &= 1 - \probc{ \,\bigvee_{i = W+1}^{M} LOS_i}{ LOS_1, \dotsc , LOS_W } \notag\\
    &= 1 - \sum_{k = 1}^{M - W} (-1)^{k+1} S_k,
    \end{align}
where we have used the inclusion-exclusion principle \cite{inclusion_exclusion} in the last step, defining $S_k$ as
\begin{equation} \label{eq:s_k}
    S_k\, = \hspace{-4mm}   \sum_{ \substack{ {\mathcal{A}\subseteq \{W+1, \dotsc, M\}} \\
    {|\mathcal{A}| = k} } } \hspace{-4mm} \probc{ \,\bigwedge_{i\in \mathcal{A}} LOS_i}{ LOS_1, \dotsc , LOS_W }.
\end{equation}

To compute the expression in \eqref{eq:nlos_cond_los}, we use \eqref{eq:prob_los} to calculate the probability $\prob{\, \bigwedge_{j \in \mathcal{J}}LOS_j } = e^{ -\esp{ K_{\mathcal{J}} } }$ for some arbitrary set of indexes $\mathcal{J}$. We can evaluate this using the formula for the expected value of $K_{\mathcal{J}}$, which was detailed in \eqref{eq:e_k_j}. Therefore, we can calculate the probabilities corresponding to the terms in the summation $S_k$ \eqref{eq:s_k} as follows:
\begin{equation}
    \probc{\, \bigwedge_{i\in \mathcal{A}} LOS_i}{ LOS_1, \dotsc , LOS_W } =\frac{\prob{\, \bigwedge_{i\in \mathcal{A}\cup\{1,\dotsc,W\}} LOS_i }}{\prob{\, \bigwedge_{i\in \{1,\dotsc,W\}} LOS_i }}
    =\frac{e^{ -\esp{ K_{\mathcal{A}\cup\{1,\dotsc,W\}} } }}{e^{ -\esp{ K_{\{1,\dotsc,W\}} } }}.
\end{equation}

%% file: Sections/implementation.tex
\section{Implementation} \label{sec:implementation}

A brief computational analysis of expression \eqref{eq:nlos_cond_los}, which is required to compute the probability given by \eqref{eq:y_conditioned_to_data}, may shed some light on the necessity to approximate these calculations. In the worst-case scenario in terms of complexity, we have $M-W = M$ (that is, $W = 0$), which implies the maximum number of terms in the summation in \eqref{eq:nlos_cond_los}. In this case, the number of terms will be the number of subsets we can obtain from $M$ elements, that is, $2^{M}$. Thus, for $N$ given data points $\{x^{(i)}\}$, a simple asymptotic lower bound on the computational complexity of \eqref{eq:y_conditioned_to_data} can be given by $\Omega(2^N)$ in the worst-case, which makes the computation of the Bayesian predictor unfeasible even for a relatively small number of data points.

\subsection{Sufficient Condition for MAP Prediction} \label{sec:bonferroni_inequalities} 

In this section, we will use the Bonferroni inequalities \cite{comtet1974bonferroni} to compute the Bayesian prediction without having to calculate all the terms in the summation in \eqref{eq:nlos_cond_los}. Recall the expression of the \ac{map} estimator \eqref{eq:map} and the complementary property of the desired probabilities. We can expand the first inequality of the predictor as before to get an equivalent relation:
\begin{align*}
    \probc{y=1}{\{x^{(i)} = d^{(i)}\}} > \frac{1}{2} \implies \prob{ y=1 , \{x^{(i)} = d^{(i)}\}} > \frac{1}{2}\prob{\{x^{(i)} = d^{(i)}\} }.
\end{align*}
Again, renumbering the data points $\{x^{(i)}\}$ so that we have the first $W$ points in LOS and the last $N-W$ in NLOS, and developing as in the previous sections (see \eqref{eq:nlos_to_cond} for analogous procedure), we can write the previous inequality as
\begin{equation*}
    \probc{NLOS_{W+1}, \dotsc, NLOS_{N}}{LOS_y, LOS_1, \dotsc, LOS_W} >\frac{1}{2} \frac{\prob{ x^{(1)} = d^{(1)}, \dotsc , x^{(N)} = d^{(N)} }}{\prob{LOS_y, LOS_1, \dotsc, LOS_W}}.
\end{equation*}
Using the probability of the complement of the left-hand side of the previous inequality yields
\begin{equation} \label{eq:los_cond_los}
    P = \probc{ \bigvee_{i = W+1}^{N} LOS_i}{ LOS_1,\dotsc, LOS_W, LOS_y } <  1-\frac{1}{2}\frac{\prob{ x^{(1)} = d^{(1)},\dotsc, x^{(N)} = d^{(N)} }}{\prob{ LOS_1, \dotsc , LOS_W, LOS_y }}.
\end{equation}
Expanding the first term with the inclusion-exclusion principle, as we have done before in \eqref{eq:nlos_cond_los}, for each $k$ such that $0 \leq k \leq N-W$ we can consider the sum $S_k$, defined previously in \eqref{eq:s_k}. Using Bonferroni's inequalities \cite{comtet1974bonferroni}, we get
%\begin{multline}
%    \sum_{k = 1}^{r-1} (-1)^{k-1} S_k \leq \\
%    {} \probc{ \bigvee_{i = W+1}^{N} LOS_i}{ LOS_1,\dotsc, LOS_W, LOS_y } \leq \\
%    {}\sum_{k = 1}^r (-1)^{k-1} S_k,
%\end{multline}
\begin{equation}
    \sum_{k = 1}^{r-1} (-1)^{k-1} S_k \leq P \leq \sum_{k = 1}^r (-1)^{k-1} S_k,
\end{equation}
if $r$ is odd, and the reversed inequalities if $r$ is even. Thus, for $r$ odd, if the inequality holds:
\begin{equation*}
    \sum_{k = 1}^r (-1)^{k-1} S_k <  1-\frac{1}{2}\frac{\prob{ x^{(1)} = d^{(1)},\dotsc, x^{(N)} = d^{(N)} }}{\prob{ LOS_1, \dotsc , LOS_W, LOS_y }},
\end{equation*}
we can conclude that $\hat{y}_{\mathrm{MAP}} = 1$; and similarly with the reversed inequality, for $r$ even, $\hat{y}_{\mathrm{MAP}} = 0$. When applicable, the derived bounds allow us to assure that the probability of the Bayesian estimator will be above or below $\frac{1}{2}$ without having to calculate the probabilities in expression \eqref{eq:los_cond_los} exactly.

\subsection{Approximated MAP Using Nearest Neighbors} \label{sec:nearest_angular_neighbors}
In this section, we will present a method to approximate the Bayesian predictor, that can reduce the time complexity of the algorithm from $\Omega(2^N)$ to $\Omega(N + k\log(N) + 2^k)$. The following approximation relies on the fact that 
\begin{equation}\label{eq:16}
    \probc{x^{(1)} = d^{(1)}}{x^{(2)} = d^{(2)}} \approx \prob{x^{(1)} = d^{(1)}},
\end{equation}
for any pair of points $x^{(1)}$, $x^{(2)}$ taking values $d^{(1)}$, $d^{(2)}$, which are sufficiently distant in terms of their angular (azimuthal) coordinate (see the \hyperref[app:validity]{Appendix} for further analysis on the validity of this approximation). This approach also holds for more than a couple of points. We can approximate the probabilities appearing in the MAP predictor as
\begin{equation*}
    \probc{ y = 1 }{ \{x^{(i)} = d^{(i)}\}_{i=1}^{N} }  \approx \probc{ y = 1 }{ \{x^{(i)} = d^{(i)}\}_{i \in \mathcal{I}} }, %\quad  \mathcal{I} \subseteq \{1 ,\dotsc, N\};
\end{equation*}
being $\{x^{(i)}\}_{i \in \mathcal{I}} $ a subset of data points that are near in angle to the point to predict, $y$. As we restrict the number of points in $\{x^{(i)}\}_{i \in \mathcal{I}}$, the value of $M-W$ in \eqref{eq:nlos_cond_los} becomes smaller, thus decreasing the exponent of the number of summands to evaluate, which is $2^{M-W}$.
We have to note that a naive approach, which would be to consider all the points that lie in an angular slice centered in the point that we want to predict, is ill-advised. If too many points fall in this slice, the algorithm will still be exponential in time. Instead of considering all the points in the slice, we can take into account only the $k$ nearest angular neighbors. This improves the approximation in the case that no points lie in the slice. Choosing the minimum $k$ values of the angle that the $N$ data points form with $y$ can be done in $\Omega(N + k\log (N))$ time\cite{cormen2009introduction}. Afterward, performing the Bayesian estimation based on this set takes $\Omega(2^k)$ operations. Thus, the total time complexity of estimating in $y$ is $\Omega(N + k\log (N) + 2^k)$. We will call this method of approximating the MAP estimation $k$N-MAP.

%% file: Sections/results_alternativo.tex
\begin{figure*}[t]
    \vspace{-0.1cm}
    %\centering
     \hspace{1.0cm}
     \subfloat{\centering \includegraphics[trim={3.6cm 9.45cm 3.2cm 10.cm},clip,width=0.43\columnwidth]{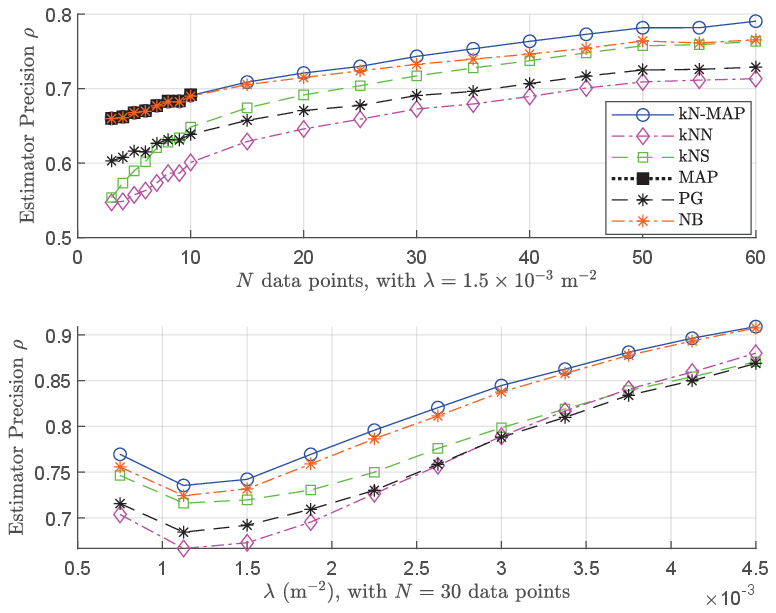}}
    \hspace{1.5cm}
    \subfloat{\centering \includegraphics[trim={3.6cm 9.45cm 3.2cm 10.0cm},clip,width=0.4\columnwidth]{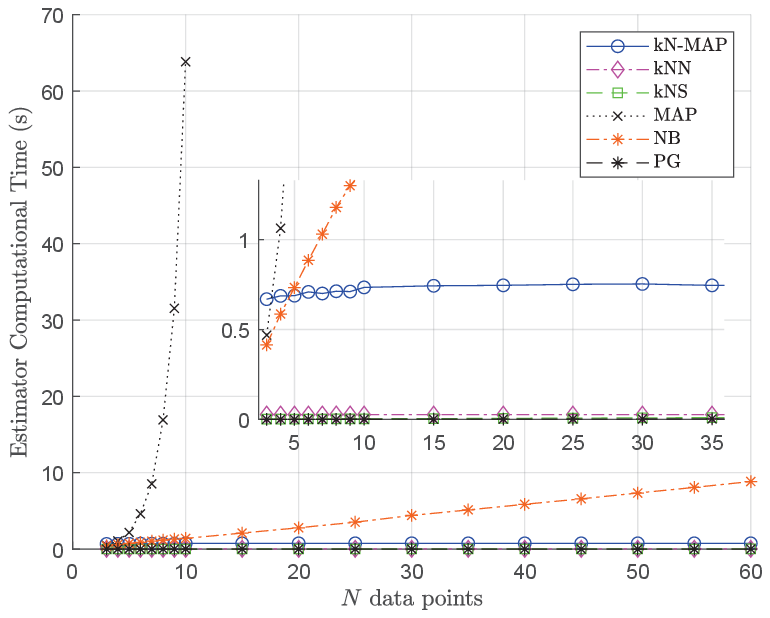}}
    \vspace{-0.1cm}
    \caption{Result of $10^3$ Monte-Carlo simulations for each amount of $N$ data points (or density of blockages $\lambda$), calculating the precision and computational time of the MAP, $k$N-MAP ($k = 3$), $k$NN ($k=3$), $k$NS ($k=1$), PG (standard deviation for the Gaussian window $\sigma=1.8$), and NB. The simulation scenario parameters were $R = 150$ m and $L_{\max} = 20$ m. For $N$ variable, the density blockages was set to $\lambda = 0.0015$ $\textrm{m}^{-2}$. The subplot on the bottom left shows $\rho$ for different values of $\lambda$, with $N=30$, to illustrate that $\lambda$ around $0.001125$ $\textrm{m}^{-2}$ is in the worst-case scenario region.}
    \label{fig:MAPvsAPX}
    \vspace{-0.7cm}
\end{figure*}

\begin{figure*}[t]
     \hspace{1cm}
    \subfloat {\includegraphics[trim={3.6cm 10.7cm 3.2cm 10.2cm},clip,width=0.44\columnwidth]{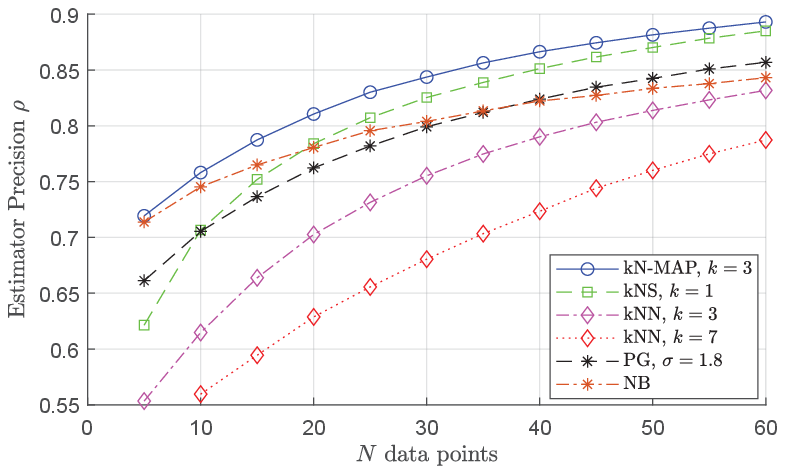}}
   \hspace{2cm}
    \subfloat{ \vspace{-2cm}\includegraphics[trim={0cm 0cm 0.2cm 0.2cm},clip, width=.26\columnwidth]{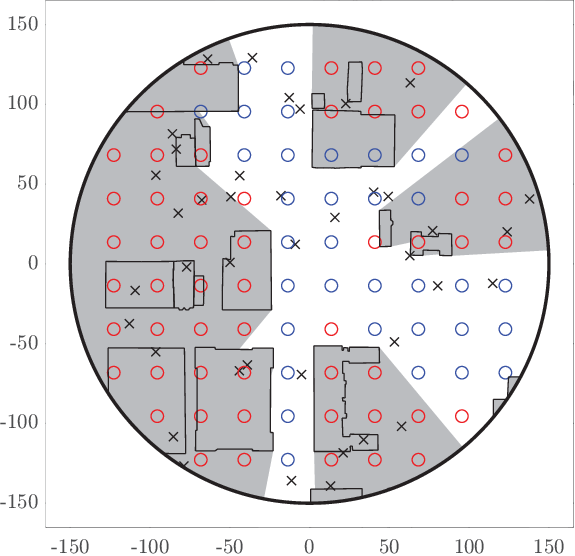}}
    \caption{Result of $10^4$ Monte-Carlo simulations using a real layout from Chicago (simulation example with layout on the right). The simulation example has blue circles for $\hat y = LOS$ and red circles for $\hat y = NLOS$. The simulation setup parameters were $R = 150$ m,  $L_{\max} = 28.6$ m ($\frac{1}{4}$ of the average perimeter of Chicago buildings), and $\lambda = 8.77\times 10^{-4}$ $\textrm{m}^{-2}$ (obtained assuming 2 obstacles per building).} 
    \label{fig:estimator_precision_chicago}
     \vspace{-0.4cm}
\end{figure*}

\vspace{-0.2cm}
\section{Results} \label{sec:results}

In this section, the performance of the proposed Bayesian estimators (MAP and $k$N-MAP) is evaluated. For comparison purposes, we also consider four other estimators:\\
(i) The \ac{knn} estimator \cite{duda} is a classifier that labels each prediction point based on the majority of the labels of the $k$ data points at minimum Euclidean distance.
\\
(ii) The \ac{kns} estimator is a novel improved version of \ac{knn} based on the geometry of the problem. It associates a segment to each data point. If the point is in LOS, the segment goes from the origin to that point and it is surely in LOS. Analogously, if the data point is in NLOS, the radial segment from the point to the circle is associated to that point and it is surely in NLOS. Accordingly, when estimating the state of a point, the majority of the $k$ nearest segments in distance determines whether that point is in LOS or not.\\
(iii) The \ac{gaussian} classifier \cite{duda} assigns a weight to each LOS/NLOS measurement. The weight is computed with a Gaussian function that depends on the distance between the measurement and prediction points and, therefore, does not require costly computations. The prediction is LOS if the summation of the weights corresponding to the LOS measurements is higher than the summation of the weights assigned to the NLOS measurements, and viceversa.\\
(iv) Finally, the \ac{naive} estimator \cite{duda} also applies Bayes' criterion but with the simplifying assumption that the measurements are independent of each other, but not with the NLOS/LOS state to be estimated at the point of interest. 

Initially, we consider randomly generated scenarios, where the lengths and orientations of the blocking elements are modeled as random with uniform distributions $L \sim \mathcal{U}(0, L_{\max})$ and $\Theta  \sim \mathcal{U}(0, \pi)$. $N$ data points are drawn from a uniform distribution on the disk. To measure the performance, we take a  uniform grid with $N_{\mathrm{est}} = 88$ points to estimate. As performance metric, we use a precision rate $\rho = \frac{N_{\mathrm{est}}-N_{\mathrm{miss}}}{N_{\mathrm{est}}}$, where $N_{\mathrm{miss}}$ is the number of miss-classified points.

\autoref{fig:MAPvsAPX} contains a performance and computational comparison of the estimators. Due to the exponential complexity of MAP, which is confirmed in the simulations, we restrict its simulations to $N \leq 10$. We observe no significant difference in precision between MAP and its approximation $k$N-MAP (see the \hyperref[app:validity]{Appendix} for an explanation). In addition to being the only feasible option, it is reasonable to use $k$N-MAP for MAP estimation given larger data sets. Regarding the other estimators, we observe that \ac{gaussian} performs better than \ac{knn} but worse than \ac{kns} for $N\geq 10$. This is because, for a sufficient set of data points, \ac{kns} captures well the geometry of the problem, while \ac{gaussian} and \ac{knn} only take into account Euclidean distances\footnote{Two points at the same Euclidean distance from the prediction point are given the same importance by \ac{gaussian} and \ac{knn}. However, note that a data point in NLOS between the BS and the prediction point is more informative than, for example, a data point in NLOS at the same distance but beyond the prediction point. This is why \ac{gaussian} and \ac{knn} do not capture the problem geometry well.}. On the other hand, the assumption taken by the \ac{naive} may be inaccurate, particularly when the data points are close. As the measurement area is fixed, this is why the performance of \ac{naive} degrades with respect to $k$N-MAP as $N$ increases, despite having equal performance for low values of $N$. 

The computational costs of the estimators, shown in \autoref{fig:MAPvsAPX} right, behave as expected. The cost of \ac{naive}, which requires calculating probabilities through integration, increases linearly with the number of data points and not exponentially as in MAP. For $N \geq 5$, the cost of \ac{naive} is higher than for $k$N-MAP and, in addition, it always performs worse. On the other hand, the $k$N-MAP computational time is considerably larger than for \ac{gaussian}, \ac{knn}, and \ac{kns}, as the constant operations performed in the Bayesian estimation \eqref{eq:e_k_j} are costly. Simulation results show analogous trends to those in \autoref{fig:MAPvsAPX} for different values of $\lambda$, but they are not included here for the sake of brevity.

Finally, \autoref{fig:estimator_precision_chicago} presents the results using a real layout of obstacles taken from a pruned version of downtown Chicago \cite{measurementsKulkarni}. They confirm the insights gained from the randomly generated scenarios: $k$N-MAP performs better than the other estimators; although, with the exception of the \ac{naive} approach, the difference reduces as $N$ increases. As already observed, \ac{naive} behaves as $k$N-MAP when the measurements are really independent (as happens for low $N$) but not when such assumption does not hold. On the other hand, \ac{kns} captures well the geometry of the problem for $N$ high enough, approaching the performance of $k$N-MAP as $N$ increases, but with significantly lower complexity than \ac{naive} and $k$N-MAP.

%% file: Sections/conclusion.tex
\section{Conclusions and Future Work}
\label{sec:conclusion}
We have analyzed the problem of estimating LOS regions for \ac{mmw} cellular urban systems. The scenario consists in a single \ac{bs} and a set of random blocking elements with a given spatial density. A Bayesian predictor is provided in order to determine whether a point is in LOS or not.

A set of methods is provided to alleviate its computational cost. First, it is proven that unnecessary sums can be avoided without changing the output of the MAP estimation. Second, calculated approximations are made, reducing the computation time from exponential to quasi-linear. Their relative error is discussed in the \hyperref[app:validity]{Appendix} and the performance is evaluated through simulations. These confirm the previously calculated computational times and shed light on the cost and performance of the proposed approaches to estimate the LOS zones. No performance loss is observed between MAP and our proposed approximation $k$N-MAP. It is also shown that $k$N-MAP performs better than our improved nearest neighbor estimator \ac{kns}, which in turn outperforms the \ac{knn}, \ac{gaussian}, and \ac{naive} methods for a sufficiently high number of data points.

The proposed procedures could prove of use: (i) When building maps are not available and no prior BS exists: in this case, the operator can try different BS positions and estimate the coverage map for each one to decide the best location. (ii) When building maps are available but they are inaccurate or out-of-date. In this case, predicting LOS/NLOS maps using updated measurements taken by UTs may be more beneficial than the predictions based on building maps. These estimated LOS/NLOS maps can be helpful, for example, to predict handovers, to know for how long moving UTs will be crossing “shadow” regions (and anticipate the impact on streaming services), or to decide if more BSs need to be deployed. Note that, if both types of information (imperfect building maps and LOS/NLOS measurements from UTs) are available, they could be combined statistically with different weights according to their information accuracy. This, however, has not been not considered in this letter and is left as future work.

%% file: Sections/appendix.tex
\appendix[Validity of the Approximation] \label{app:validity}
A short discussion of the approximation made in \eqref{eq:16} is presented here. Consider two points $x^{(1)}$, $x^{(2)}$. We have seen how the probability $\probc{x^{(1)} = d^{(1)}}{x^{(2)} = d^{(2)}}$ can be reduced to a calculation of the form $\probc{LOS_1}{LOS_2} = e^{-\esp{K_{1,2}}+\esp{K_2}}$ \footnote{
    As an example consider $\probc{LOS_1}{NLOS_2} = \frac{\prob{LOS_1 \wedge NLOS_2}}{\prob{NLOS_2}} = \frac{\probc{NLOS_2}{LOS_1}}{1-\prob{LOS_2}}\prob{LOS_1} = \frac{1 - \probc{LOS_2}{LOS_1}}{1-\prob{LOS_2}}\prob{LOS_1}$
}, interchanging indices if necessary, where $$\esp{K_{1,2}} = \int_l\int_\theta \lambda f_L(l) f_{\Theta}(\theta) A_{S_{1_{l\theta}}\cup S_{2_{l\theta}}}.$$ If the points $x^{(1)}$ and $x^{(2)}$ are sufficiently separated in angle, for any given orientation and length $\theta, l$, the overlap of the quadrilaterals $S_{1_{l\theta}}$ and $S_{2_{l\theta}}$ can be neglected, thus having $A_{S_{1_{l\theta}}\cup S_{2_{l\theta}}} \approx A_{S_{1_{l\theta}}} + A_{S_{2_{l\theta}}}$, hence 
\begin{align*}
    \esp{K_{1,2}} &= \int_l\int_\theta \lambda f_L(l) f_{\Theta}(\theta) A_{S_{1_{l\theta}}\cup S_{2_{l\theta}}} \\
    & \approx \int_l\int_\theta \lambda f_L(l) f_{\Theta}(\theta) A_{S_{1_{l\theta}}} \hspace{-0.05cm} + \hspace{-0.05cm} \int_l\int_\theta \lambda f_L(l) f_{\Theta}(\theta) A_{S_{2_{l\theta}}} \\
    &= \esp{K_1} + \esp{K_2}.
\end{align*}
Therefore, 
\begin{equation*}
    \probc{LOS_1}{ LOS_2} = e^{-\esp{K_{1,2}}+\esp{K_2}} \approx e^{-\esp{K_1}-\esp{K_2}+\esp{K_2}} = e^{-\esp{K_1}} =  \prob{LOS_1}.
\end{equation*}
This expression provides the approximation made in \eqref{eq:16}. 

\begin{figure}[t]
    \centering
    \includegraphics[trim={2.6cm 9.4cm 2.6cm 10cm},clip,width=0.45\columnwidth]{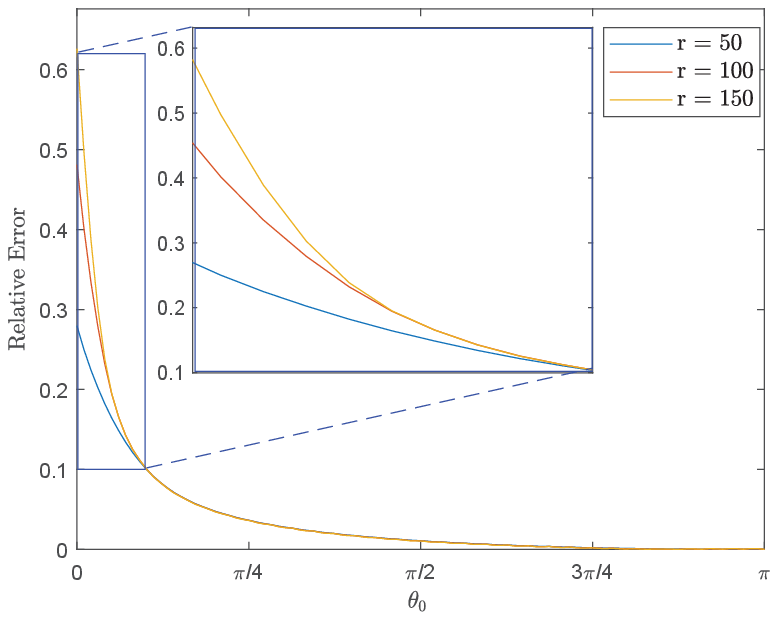}
    \caption{Relative error depending on the angle between points.}
    \label{fig:relative_error}
\end{figure}

To illustrate the previous reasoning, we have computed the relative error from this approximation when both points are in LOS (see \autoref{fig:relative_error}). For simplicity, we have considered both points to be at the same distance from the \ac{bs}. The points are at an angle difference $\theta_0$. The random modeling of the blocking elements is the same as in \autoref{sec:results}, \autoref{fig:MAPvsAPX}, i.e. $f_{L}(l)$ and $f_{\Theta}(\theta)$ are both uniform PDFs.
As we can see, the relative error decays quickly and, for points that are at an angular distance larger than $\pi/4$, it is below $0.05$. We also notice that the error is always lower for points that are closer to the \ac{bs}. Intuitively, this is explained as follows: points closer to the \ac{bs} are always less likely to be blocked than more distant points; therefore, there is less uncertainty with respect to their prediction.